\documentclass[mathleft,cite,graphics,graphicx,psfig,amssymb,landscape]{an}
\usepackage{graphicx}
\usepackage{amsfonts}
\usepackage{natbib}
\usepackage{times}
\usepackage{lscape}
\begin{document}
\Pagespan{1}{}%
\Yearpublication{2010}%
\Yearsubmission{2010}%
\Month{XX}%
\Volume{XX}%
\Issue{XX}%
\DOI{DOI}%

\title{Transformations between 2MASS, SDSS and BVI photometric systems for late--type giants}

\author{E. Yaz\inst{1}\fnmsep\thanks{Corresponding author: \email{esmayaz@istanbul.edu.tr}\newline} 
\and S. Bilir\inst{1}
\and S. Karaali\inst{1,2}
\and S. Ak\inst{1}
\and B. Co\c skuno\u glu\inst{1}
\and A. Cabrera-Lavers \inst{3,4}
}

\institute{Istanbul University, Faculty of Sciences, Department of Astronomy and 
Space Sciences, 34119 Istanbul, Turkey
\and
Beykent University, Faculty of Science and Letters, Department of Mathematics  
and Computer, Beykent, 34398, Istanbul, Turkey
\and
Instituto de Astrof\'{\i}sica de Canarias, E-38205 La Laguna, Tenerife, Spain
\and
GTC Project Office, E-38205 La Laguna, Tenerife, Spain
}
\date{} 

\received{}
\accepted{}
\publonline{later} 

\abstract{We present colour transformations from Two Micron All Sky Survey (2MASS) photometric 
system to Johnson–-Cousins system and to Sloan Digital Sky Survey (SDSS) system for 
late–-type giants and vice versa. The giant star sample was formed using surface gravity 
constraints ($2 < \log g \leq 3$) to Cayrel de Strobel et al.\rq s (2001) spectroscopic catalogue. 
2MASS, SDSS and Johnson–-Cousins photometric data was taken from \cite{Cu03}, \cite{Ofek08} 
and \cite{vanLeeuwen2007}, respectively. The final sample was refined applying the following steps: 
(1) the data were dereddened, (2) the sample stars selected are of the highest photometric quality. 
We give two--colour dependent transformations as a function of metallicity as well as independent 
of metallicity. The transformations provide absolute magnitudes and distance determinations which 
can be used in space density evaluations at relatively short distances where some or all of the 
SDSS magnitudes of late--type giants are saturated.}

\maketitle

\section{Introduction}
The most widely used sky surveys are the Sloan Digital Sky Survey 
\citep[SDSS;][]{York00} and the Two Micron All Sky Survey \citep[2MASS;][]{2mass06}.  
SDSS is the largest photometric and spectroscopic survey in optical wavelengths,
whereas, 2MASS has imaged the sky across infrared wavelengths. Another astrometrically and 
photometrically important survey is Hipparcos \citep{ESA97}, which was reduced recently by \cite{vanLeeuwen2007}.

SDSS obtains images almost simultaneously in five broad--bands ($u$, $g$, $r$, $i$ 
and $z$) centred at 3540, 4760, 6280, 7690 and 9250\AA, respectively \citep{Fukugita96, 
Gunn98, Hogg01, Smith02}. The photometric pipeline \citep{Lupton01} detects the objects, 
matches the data from five filters and measures instrumental fluxes, positions and shape 
parameters. The magnitudes derived from fitting a point spread function (PSF) 
are accurate to about 2 per cent in $g$, $r$ and $i$, and 3--5 per 
cent in $u$ and $z$ for bright sources ($<$ 20 mag) point sources. Data Release 5 (DR5) 
is almost 95 per cent complete for point sources to ($u$, $g$, $r$, $i$, $z$)= 
(22, 22.2, 22.2, 21.3, 20.5). The median FWHM of the PSFs is about 
1.5 arcsec \citep{Abazajian04}. The data are saturated at about 14 mag in $g$, 
$r$ and $i$, and about 12 mag in $u$ and $z$ \citep[see][]{Chonis08}.

The near-infrared (NIR) $JHK_{s}$ photometric data were taken from the 
digital Two Micron All-sky Survey\footnote{http://www.ipac.caltech.edu/2MASS/} 
(2MASS). It provides the most complete database of galactic point sources available 
up to date. During the development of this survey, two highly automated 1.3--m 
telescopes were used: one at Mt. Hopkins, Arizona to observe the Northern sky, 
and the other at Cerro Tololo Observatory in Chile to complete the survey{\rq}s 
Southern half. Observations cover 99.998 per cent \citep{2mass06} of the sky with 
simultaneous detections in $J$ (1.25 $\mu m$), $H$ (1.65 $\mu m$) and $K_{s}$ 
(2.17 $\mu m$) bands up to the limiting magnitudes of 15.8, 15.1 and 14.3, 
respectively. Nowadays, 2MASS is probably the reference study for the 
rest of galactic surveys, due to both full coverage of the sky which it provides 
and its intrinsic photometric and astrometric accuracies. Recent surveys, such as 
UKIDSS \citep{Lucas2008}, use 2MASS as reference catalogue, therefore a 
transformation between optical and 2MASS magnitudes is of interest in many 
different topics. The photometric uncertainty of the data is less than 0.155 at 
$K_{s} \sim 16.5$ magnitude which is the photometric completeness of 2MASS 
for stars with $|b| > 25^{o}$ \citep{2mass06}. Calibration offsets between any 
two points in the sky are less than 0.02 mag. The passband profiles for 
Johnson–-Cousins, SDSS and 2MASS photometric systems are given in Fig. 1. 

\begin{figure}
\begin{center}
\includegraphics[angle=0, width=80mm, height=98.7mm]{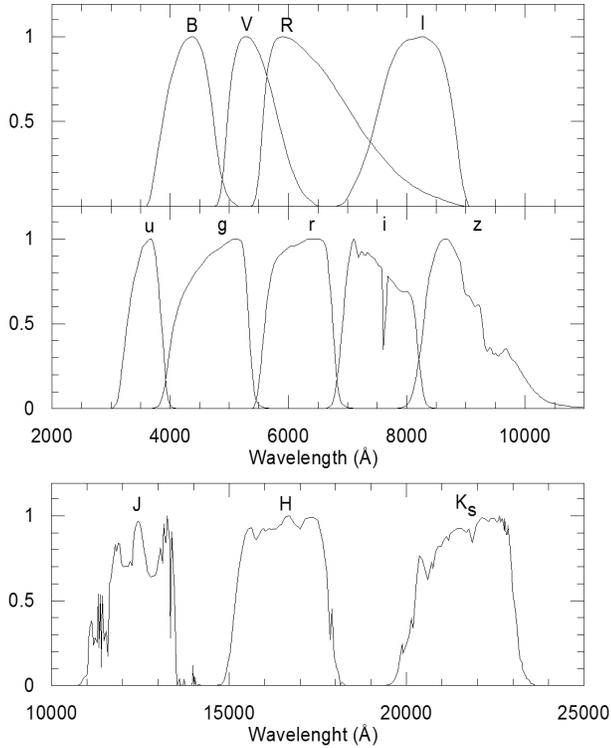}
\caption[] {Normalized passbands of the Johnson-Cousins\rq $BVRI$
filters (upper panel), the {\em SDSS} $ugriz$ filters (middle
panel), and the {\em 2MASS} filters (lower panel).}
\end{center}
\end{figure}

It has been custom to derive transformations between a newly defined photometric system 
and those that are more traditional (such as the Johnson--Cousins $UBVRI$ system). 
A number of transformations between $u'g'r'i'z'$, $ugriz$ and $UBVR_{c}I_{c}$ exist. The 
$u'g'r'i'z'$ system is referred to the similar filter system used on the 0.5--m Photometric 
Calibration Telescope at SDSS. It should be noted that there are differences between the 
$u'g'r'i'z'$ and $ugriz$ systems. These are discussed in \cite{Tucker06}, \cite{Davenport07} 
and \cite{Smith07}. In this paper we are concerned with transformations to and from $ugriz$ 
system of the 2.5--m. The first transformations derived between the SDSS $u'g'r'i'z'$ system 
and the Johnson--Cousins\rq photometric system were based on the observations in  
$u'$, $g'$, $r'$, $i'$ and $z'$ filters \citep{Smith02}. 

An improved set of transformations between the observations obtained in $u'g'r'$ 
filters at the Isaac Newton Telescope (INT) at La Palma, Spain, and the \cite{Landolt92} 
$UBV$ standards is derived by \cite{Karaali05}. The INT filters were designed to reproduce 
the SDSS system. \cite{Karaali05} presented for the first time transformation equations 
depending on two colors.

\cite{Rodgers06} considered two–-color or quadratic forms in their transformation 
equations. \cite{Jordi06} used SDSS DR4 and $BVRI$ photometry taken from different 
sources and derived population (and metallicity) dependent transformation equations 
between SDSS and $UBVRI$ systems. A recent work is by \cite{Chonis08} who used transformations 
from SDSS $ugriz$ to $UBVRI$ not depending on luminosity class or metallicity to 
determine CCD zero--points. Finally, we refer to our most recent paper where 
transformations between SDSS (and 2MASS) and $BVRI$ photometric systems for 
dwarfs are given \citep{Bilir08}. 

The first transformations between 2MASS and other photometric systems are those 
of \cite{Walkowicz04} and \cite{West05} who determined the level of magnetic 
activity in M and L dwarfs. The aim of \cite{Davenport07} in deriving equations 
between 2MASS and other photometric systems was to estimate the absolute magnitudes 
of cool stars. \cite{Covey08} considered the $ugrizJHK_{s}$ stellar locus and showed how it 
can be used to identify objects with unusual colors. A recent study by \cite{Straizys09} 
provided calibrations which can be used to obtain the color indices of 2MASS stars 
with known spectral types and luminosity classes. \cite{Straizys09} used the loci 
of giant stars with spectral types later than G5 to obtain their calibrations. 

To extend the results of \cite{Bilir08} to the giants domain is of great use, as 
it allows to derive absolute magnitudes and to produce distance determinations at 
relatively short distances where SDSS sources will be saturated. For example, red clump 
giants are now considered as well-known standard candles, as they show 
a very narrow luminosity function that constitute a compact and well-defined clump 
in a Hertzsprung-Russell diagram, particularly in the NIR. The absolute magnitude 
($M_{K}$) and intrinsic colors, ($J-K_{s}$)$_{0}$, of the red clump giants are 
well established with very small metallicity dependences \citep[see][and references 
therein]{Cabrera2008}, hence distances to these sources can be derived with confidence 
by means of a very small number of assumptions \citep{Lopez2002}. This well known 
set of NIR properties of the red clump population have not been sufficiently studied 
in the optical, where a lot of SDSS data is waiting to be exploited. In order to do this, 
a proper set of optical/NIR photometric transformations are needed and these are presented here.

In Section 2 we present the sources of our star sample and the criteria applied to 
the chosen stars. The transformation equations are given in Section 3. Finally, 
in Section 4, we discuss our results.
 
\section{Data}
The first and main source of our data was Cayrel de Strobel et al.'s (2001) spectroscopic 
catalogue. This catalogue contains a large number of stars with different population types 
and metallicities. We chose 661 stars with $2 < \log g \leq 3$ and obtained the original 
sample of late--type giants. The second source for our work is the 2MASS All--Sky Catalogue 
of Point Sources \citep{Cu03}. We matched the Cayrel de Strobel et al.\rq s (2001) spectroscopic 
catalogue with the Cutri et al.\rq s (2003) 2MASS photometric catalogue and obtained the NIR 
magnitudes for 661 stars. The NIR magnitudes of 661 stars are not as precise as their 
optical magnitudes. To select more sensitive NIR magnitudes of sample stars, we used the 
magnitude flags, labeled ``AAA'', which means the signal noise ratio is $SNR \geq 10$, 
i.e. they have the highest quality measurements. 
\begin{figure*}
\begin{center}
\includegraphics[scale=0.65, angle=0]{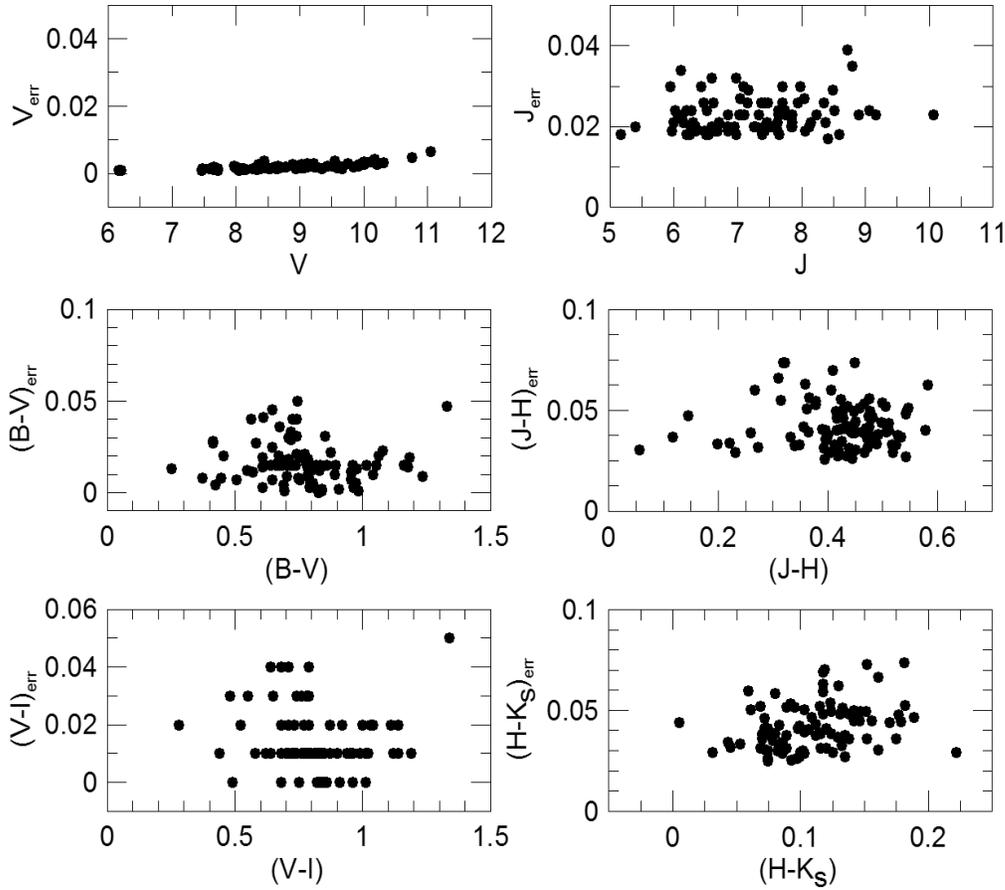}
\caption[] {The error distributions for Johnson-Cousins\rq $BVI$ and 2MASS 
$JHK_{s}$ (the errors for SDSS $gri$ magnitudes are discussed in the text).}
\end{center}
\end{figure*}

After applying this selection criterion based on the quality of the data, the total number of stars in 2MASS photometric system was 
reduced to 91. The photometric and spectroscopic data of 91 stars is given in Table 1. 

The $BVI$ magnitudes were taken from the newly reduced Hipparcos\rq \citep{vanLeeuwen2007} 
catalogue. The same catalogue also offers trigonometric parallaxes which provides accurate 
distance determination. Finally, the $g$, $r$ and $i$ magnitudes were taken from the Ofek's 
(2008) catalogue. Ofek (2008) calculated the synthetic $griz$ magnitudes of bright Tycho stars, 
where SDSS magnitudes are actually saturated. Out of 91 stars only 82 had $g$, $r$ and $i$ 
magnitudes, due to the shallowness of 2MASS and deepness of the SDSS photometries. Thus, 
transformation equations between $BVI$ and 2MASS is carried out for a sample of 91 stars, 
whereas the ones between SDSS and 2MASS is limited with 82 stars. 

The errors for the colors and magnitudes for {\em BVI} and 2MASS photometries are 
given in Fig. 2. For SDSS magnitudes, \cite{Ofek08} states that the typical 
errors for a single star are 0.12, 0.12, 0.10 and 0.08 mag for $g$, $r$, $i$ 
and $z$ bands, respectively, but they reduce to about 0.04, 0.03, 0.02 and 
0.02 mag for a sample of about 10 stars. 

The $E(B-V)$ color excesses of stars have been evaluated in two steps. 
First, we used the maps of \cite{Schlegel98} and evaluated a $E_{\infty}(B-V)$ 
color excess for each star. We then reduced them using the following procedure \citep{Bahcall80}:

\begin{equation}
A_{d}(b)=A_{\infty}(b)\Biggl[1-\exp\Biggl(\frac{-\mid
d\sin(b)\mid}{H}\Biggr)\Biggr].
\end{equation}
Here, $b$ and $d$ are the galactic latitude and distance of the star, respectively. 
$H$ is the scaleheight for the interstellar dust which is adopted as 125 pc 
\citep{Marshall06} and $A_{\infty}(b)$ and $A_{d}(b)$ are the total absorptions for the
model and for the distance to the star, respectively. $A_{\infty}(b)$ can be evaluated 
by means of the following equation:
\begin{equation}
A_{\infty}(b)=3.1E_{\infty}(B-V).
\end{equation}
$E_{\infty}(B-V)$ is the color excess for the model taken from the NASA Extragalactic
Database\footnote{http://nedwww.ipac.caltech.edu/forms/calculator.html}. Then, $E_{d}(B-V)$, 
i.e. the color excess for the corresponding star at the distance $d$, can be evaluated by 
Eq. (3) adopted for distance $d$,

\begin{equation}
E_{d}(B-V)=A_{d}(b)~/~3.1.
\end{equation}

\setcounter{table}{0}
\begin{landscape}
\begin{table*}
\setlength{\tabcolsep}{1.4pt}
\
\caption{Johnson-Cousins, {\em SDSS} and {\em 2MASS} magnitudes, coordinates and parallax of
the sample stars (91 total stars). The columns give: (1) ID, (2) Hipparcos number, (3)-(5) surface gravity, 
metal abundance and their references (6) parallax, (7) relative parallax error, (8) distance (9) $E_{d}(B-V)$ reduced color excess,
(10) $V$ apparent magnitude, (11) and (12) $B-V$ and $V-I$ color indices, (13) $g$ apparent 
magnitude, (14) and (15) $g-r$, $r-i$ color indices, (16) $J$ apparent magnitude, (17) and (18) 
$J-H$ and $H-K_{s}$ color indices.}
{\tiny
\begin{tabular}{crcccccccccccccccc}
\hline
ID & Hip & $\log g$ & $[M/H]$ & Ref. & $\pi$ & $\sigma_{\pi}/{\pi}$ & $d$ & $E_{d}(B-V)$ & $V$ & $B-V$ & $V-I$ & $g$ & $g-r$ & $r-i$ & $J$ & $J-H$ & $H-K_{s}$ \\
   &        &     & (dex)   &      & (mas) &  & (pc) & (mag) & (mag) & (mag) & (mag) & (mag) & (mag) & (mag) & (mag) & (mag) & (mag) \\
\hline  
1  &   434& 2.80 & -1.39& 1 & $2.58\pm1.10$ & 0.43 & 388 & 0.016 & $9.040\pm0.002$ & $0.692\pm0.001$ & $0.75\pm0.01$ & 9.330 & 0.480 & 0.150 & $7.704\pm0.030$ & $0.356\pm0.042$ & $0.099\pm0.042$ \\
2  &   447& 2.50 & -0.45& 2 & $2.99\pm1.16$ & 0.39 & 334 & 0.074 & $8.260\pm0.002$ & $1.052\pm0.015$ & $1.11\pm0.02$ & 8.790 & 0.730 & 0.270 & $6.540\pm0.019$ & $0.434\pm0.028$ & $0.074\pm0.026$ \\
3  &   484& 2.60 & -1.19& 3 & $0.37\pm1.33$ & 3.59 & 540*& 0.021 & $9.660\pm0.002$ & $0.787\pm0.016$ & $0.82\pm0.01$ &10.040 & 0.660 & 0.180 & $8.057\pm0.019$ & $0.445\pm0.039$ & $0.099\pm0.041$ \\
4  &   999& 3.00 & -0.40& 4 &$24.38\pm0.95$ & 0.04 &  41 & 0.011 & $8.440\pm0.004$ & $0.739\pm0.015$ & $0.79\pm0.01$ & 8.810 & 0.670 & 0.250 & $6.847\pm0.023$ & $0.467\pm0.029$ & $0.135\pm0.027$ \\
5  &  1298& 2.60 & -1.17& 1 & $1.06\pm1.67$ & 1.58 & 520*& 0.020 & $9.580\pm0.002$ & $0.710\pm0.030$ & $0.76\pm0.03$ & 9.850 & 0.480 & 0.150 & $8.147\pm0.021$ & $0.361\pm0.051$ & $0.103\pm0.051$ \\
6  &  2023& 2.50 & -0.11& 5 & $7.29\pm0.76$ & 0.10 & 137 & 0.021 & $8.000\pm0.002$ & $0.253\pm0.013$ & $0.28\pm0.02$ & 8.160 & 0.120 & 0.000 & $7.415\pm0.020$ & $0.056\pm0.030$ & $0.031\pm0.029$ \\
7  &  2413& 2.50 & -2.00& 6 & $3.15\pm0.78$ & 0.25 & 317 & 0.086 & $7.720\pm0.001$ & $0.747\pm0.008$ & $0.79\pm0.01$ &   $-$ &   $-$ &   $-$ & $6.026\pm0.024$ & $0.423\pm0.041$ & $0.129\pm0.037$ \\
8  &  2463& 2.25 & -2.10& 7 & $0.88\pm0.81$ & 0.92 & 313*& 0.019 & $8.480\pm0.002$ & $0.713\pm0.015$ & $0.77\pm0.01$ & 8.800 & 0.660 & 0.260 & $6.853\pm0.020$ & $0.462\pm0.039$ & $0.135\pm0.038$ \\
9  &  2727& 2.60 & -1.04& 3 & $3.47\pm1.24$ & 0.36 & 288 & 0.018 & $9.750\pm0.003$ & $0.735\pm0.015$ & $0.78\pm0.01$ &10.080 & 0.480 & 0.150 & $8.376\pm0.021$ & $0.365\pm0.039$ & $0.106\pm0.041$ \\
10 &  3554& 3.00 & -2.30& 7 & $5.01\pm1.32$ & 0.26 & 200 & 0.029 & $9.020\pm0.003$ & $0.745\pm0.050$ & $0.79\pm0.04$ & 9.360 & 0.640 & 0.210 & $7.424\pm0.026$ & $0.449\pm0.074$ & $0.152\pm0.073$ \\
11 &  4930& 2.80 & -0.25& 8 & $2.59\pm0.94$ & 0.36 & 386 & 0.017 & $7.990\pm0.002$ & $0.983\pm0.001$ & $0.96\pm0.01$ & 8.530 & 0.780 & 0.260 & $6.203\pm0.018$ & $0.528\pm0.038$ & $0.116\pm0.039$ \\
12 &  4960& 2.53 & -0.99& 9 & $3.18\pm0.93$ & 0.29 & 314 & 0.027 & $8.600\pm0.002$ & $0.741\pm0.031$ & $0.79\pm0.02$ & 9.050 & 0.640 & 0.210 & $7.088\pm0.030$ & $0.442\pm0.040$ & $0.071\pm0.031$ \\
13 &  5104& 2.30 & -0.93&10 & $3.19\pm0.79$ & 0.25 & 313 & 0.023 & $7.650\pm0.002$ & $0.785\pm0.010$ & $0.82\pm0.01$ & 8.090 & 0.640 & 0.210 & $6.149\pm0.021$ & $0.473\pm0.032$ & $0.089\pm0.031$ \\
14 &  5445& 2.50 & -1.56& 6 & $7.16\pm0.84$ & 0.12 & 140 & 0.021 & $7.720\pm0.002$ & $0.701\pm0.009$ & $0.76\pm0.01$ & 8.090 & 0.630 & 0.210 & $6.197\pm0.024$ & $0.438\pm0.031$ & $0.097\pm0.026$ \\
15 &  5455& 2.79 & -0.83& 9 & $7.23\pm0.86$ & 0.12 & 138 & 0.014 & $9.020\pm0.002$ & $0.803\pm0.004$ & $0.83\pm0.01$ & 9.440 & 0.660 & 0.180 & $7.459\pm0.021$ & $0.466\pm0.053$ & $0.092\pm0.053$ \\
16 &  6894& 2.30 & -1.57& 9 & $2.55\pm1.04$ & 0.41 & 392 & 0.010 & $8.930\pm0.002$ & $0.820\pm0.014$ & $0.84\pm0.01$ & 9.320 & 0.660 & 0.250 & $7.289\pm0.020$ & $0.458\pm0.051$ & $0.124\pm0.050$ \\
17 & 12028& 2.10 & -0.21&11 & $4.12\pm1.17$ & 0.28 & 243 & 0.021 & $8.150\pm0.001$ & $1.235\pm0.009$ & $1.19\pm0.01$ & 8.850 & 0.950 & 0.330 & $6.062\pm0.023$ & $0.518\pm0.029$ & $0.222\pm0.029$ \\
18 & 12353& 3.00 & -2.23&12 & $2.33\pm1.20$ & 0.52 &664* & 0.027 &$10.110\pm0.004$ & $0.782\pm0.014$ & $0.82\pm0.01$ &10.420 & 0.600 & 0.220 & $8.595\pm0.018$ & $0.416\pm0.040$ & $0.080\pm0.058$ \\
19 & 14747& 3.00 & -1.03& 7 & $2.60\pm0.74$ & 0.28 & 385 & 0.018 & $8.900\pm0.002$ & $0.840\pm0.002$ & $0.86\pm0.01$ & 9.370 & 0.670 & 0.250 & $7.332\pm0.023$ & $0.482\pm0.037$ & $0.070\pm0.036$ \\
20 & 16214& 2.11 & -1.75&13 & $4.03\pm1.00$ & 0.25 & 248 & 0.103 & $8.700\pm0.002$ & $0.786\pm0.018$ & $0.82\pm0.01$ & 9.110 & 0.670 & 0.220 & $6.975\pm0.018$ & $0.428\pm0.050$ & $0.133\pm0.051$ \\
21 & 17001& 2.60 & -2.30&14 & $3.93\pm1.53$ & 0.39 & 254 & 0.015 & $9.920\pm0.002$ & $0.647\pm0.045$ & $0.71\pm0.04$ &   $-$ &   $-$ &   $-$ & $8.497\pm0.029$ & $0.400\pm0.038$ & $0.138\pm0.036$ \\
22 & 18742& 2.61 & -2.31& 9 & $2.55\pm0.70$ & 0.27 & 392 & 0.066 & $8.780\pm0.002$ & $0.709\pm0.018$ & $0.76\pm0.01$ & 9.140 & 0.630 & 0.210 & $7.172\pm0.029$ & $0.477\pm0.049$ & $0.061\pm0.051$ \\
23 & 21473& 2.20 & -0.69&10 & $1.23\pm0.65$ & 0.53 &449* & 0.009 & $9.260\pm0.002$ & $0.907\pm0.002$ & $0.91\pm0.01$ & 9.740 & 0.720 & 0.260 & $7.588\pm0.020$ & $0.480\pm0.047$ & $0.151\pm0.049$ \\
24 & 23799& 2.20 &  0.19&15 & $4.20\pm0.76$ & 0.18 & 238 & 0.036 & $6.200\pm0.001$ & $0.444\pm0.008$ & $0.55\pm0.03$ & 6.420 & 0.360 & 0.100 & $5.167\pm0.018$ & $0.145\pm0.048$ & $0.118\pm0.048$ \\
25 & 26040& 3.00 & -1.76& 7 & $2.99\pm0.86$ & 0.29 & 334 & 0.024 & $9.430\pm0.002$ & $0.740\pm0.021$ & $0.79\pm0.01$ & 9.810 & 0.640 & 0.210 & $7.856\pm0.023$ & $0.415\pm0.050$ & $0.142\pm0.050$ \\
26 & 28887& 3.00 & -1.64&12 & $2.25\pm1.92$ & 0.85 &728* & 0.174 &$10.31\pm0.003$  & $0.676\pm0.036$ & $0.74\pm0.03$ &10.590 & 0.480 & 0.150 & $8.891\pm0.023$ & $0.359\pm0.063$ & $0.118\pm0.063$ \\
27 & 29759& 2.90 & -2.31&16 & $6.46\pm1.31$ & 0.20 & 155 & 0.054 & $8.920\pm0.002$ & $0.611\pm0.041$ & $0.68\pm0.04$ & 9.240 & 0.480 & 0.160 & $7.532\pm0.020$ & $0.377\pm0.053$ & $0.069\pm0.052$ \\
28 & 29992& 2.20 & -1.61&17 & $5.57\pm0.84$ & 0.15 & 180 & 0.065 & $8.050\pm0.001$ & $0.829\pm0.012$ & $0.85\pm0.01$ & 8.460 & 0.670 & 0.230 & $6.340\pm0.020$ & $0.534\pm0.037$ & $0.103\pm0.039$ \\
29 & 30668& 3.00 & -1.58&18 & $6.72\pm0.70$ & 0.10 & 149 & 0.063 & $8.000\pm0.002$ & $0.703\pm0.015$ & $0.76\pm0.01$ & 8.360 & 0.610 & 0.220 & $6.591\pm0.020$ & $0.321\pm0.074$ & $0.181\pm0.074$ \\
30 & 34795& 2.80 & -1.55&19 & $1.94\pm0.91$ & 0.47 & 515 & 0.101 & $8.370\pm0.002$ & $0.955\pm0.007$ & $1.00\pm0.02$ & 8.890 & 0.840 & 0.320 & $6.590\pm0.032$ & $0.547\pm0.051$ & $0.112\pm0.043$ \\
31 & 37335& 3.00 & -1.02&20 & $5.17\pm1.32$ & 0.26 & 193 & 0.022 & $9.230\pm0.002$ & $0.820\pm0.002$ & $0.84\pm0.01$ & 9.640 & 0.660 & 0.260 & $7.621\pm0.021$ & $0.437\pm0.042$ & $0.146\pm0.045$ \\
32 & 37339& 3.00 &  0.25&15 & $9.84\pm0.47$ & 0.05 & 102 & 0.016 & $6.160\pm0.001$ & $0.374\pm0.008$ & $0.44\pm0.01$ & 6.340 & 0.240 & 0.010 & $5.404\pm0.020$ & $0.117\pm0.037$ & $0.071\pm0.037$ \\
33 & 40068& 3.00 & -2.05&21 & $4.97\pm1.07$ & 0.22 & 201 & 0.015 &$10.010\pm0.003$ & $0.564\pm0.040$ & $0.64\pm0.04$ &   $-$ &   $-$ &   $-$ & $8.711\pm0.039$ & $0.378\pm0.054$ & $0.072\pm0.046$ \\
34 & 44636& 2.50 &  0.13&22 & $0.84\pm0.99$ & 1.18 &327* & 0.077 & $8.570\pm0.002$ & $1.160\pm0.015$ & $1.12\pm0.01$ & 9.190 & 0.950 & 0.330 & $6.472\pm0.026$ & $0.499\pm0.054$ & $0.182\pm0.052$ \\
35 & 44716& 3.00 & -1.02&12 & $1.84\pm1.23$ & 0.67 &217* & 0.027 & $7.680\pm0.001$ & $0.796\pm0.014$ & $0.83\pm0.01$ & 8.080 & 0.650 & 0.190 & $6.105\pm0.034$ & $0.511\pm0.043$ & $0.043\pm0.034$ \\
36 & 45845& 2.30 & -0.89&10 & $3.13\pm0.60$ & 0.19 & 319 & 0.061 & $8.410\pm0.002$ & $0.975\pm0.013$ & $1.02\pm0.01$ & 8.860 & 0.790 & 0.270 & $6.516\pm0.024$ & $0.542\pm0.048$ & $0.142\pm0.045$ \\
37 & 47139& 2.20 & -1.52& 6 & $0.96\pm0.77$ & 0.80 &292* & 0.034 & $8.330\pm0.001$ & $1.015\pm0.015$ & $0.99\pm0.01$ & 8.940 & 0.980 & 0.370 & $6.307\pm0.021$ & $0.578\pm0.040$ & $0.120\pm0.041$ \\
38 & 47599& 2.40 & -1.40&23 & $0.53\pm0.92$ & 1.74 &466* & 0.010 & $9.340\pm0.002$ & $0.644\pm0.025$ & $0.71\pm0.02$ & 9.570 & 0.480 & 0.150 & $7.860\pm0.021$ & $0.393\pm0.031$ & $0.100\pm0.029$ \\
39 & 48444& 2.20 & -2.45& 6 & $0.90\pm1.50$ & 1.67 &611* & 0.010 & $9.930\pm0.003$ & $0.835\pm0.015$ & $0.85\pm0.01$ &   $-$ &   $-$ &   $-$ & $8.355\pm0.026$ & $0.424\pm0.055$ & $0.115\pm0.052$ \\
40 & 48979& 2.10 & -0.97&10 & $0.96\pm1.26$ & 1.31 &540* & 0.048 & $9.660\pm0.002$ & $0.826\pm0.001$ & $0.85\pm0.01$ &10.090 & 0.660 & 0.250 & $8.050\pm0.027$ & $0.470\pm0.041$ & $0.089\pm0.037$ \\
41 & 49371& 2.58 & -2.02&24 & $3.62\pm1.22$ & 0.34 & 276 & 0.005 & $8.970\pm0.002$ & $0.719\pm0.033$ & $0.77\pm0.02$ &   $-$ &   $-$ &   $-$ & $7.471\pm0.026$ & $0.425\pm0.035$ & $0.103\pm0.029$ \\
42 & 49616& 3.00 & -2.23&12 & $8.75\pm0.62$ & 0.07 & 114 & 0.054 & $8.660\pm0.002$ & $0.663\pm0.016$ & $0.72\pm0.01$ & 9.020 & 0.640 & 0.210 & $7.032\pm0.023$ & $0.428\pm0.046$ & $0.177\pm0.048$ \\
43 & 50601& 2.80 &  0.10&19 & $5.12\pm0.77$ & 0.15 & 195 & 0.028 & $7.480\pm0.001$ & $0.958\pm0.011$ & $0.95\pm0.01$ & 8.020 & 0.820 & 0.270 & $5.941\pm0.030$ & $0.476\pm0.056$ & $0.147\pm0.050$ \\
44 & 50994& 2.10 &  0.19&11 & $1.72\pm1.04$ & 0.60 &515* & 0.107 & $9.560\pm0.002$ & $1.330\pm0.047$ & $1.34\pm0.05$ &10.140 & 0.950 & 0.330 & $7.375\pm0.026$ & $0.544\pm0.049$ & $0.189\pm0.047$ \\
45 & 53561& 2.60 & -0.02&22 & $2.58\pm0.79$ & 0.31 & 388 & 0.043 & $8.450\pm0.002$ & $0.960\pm0.015$ & $0.95\pm0.01$ & 8.900 & 0.730 & 0.270 & $6.712\pm0.021$ & $0.474\pm0.047$ & $0.131\pm0.048$ \\
46 & 53810& 2.10 &  0.01&11 & $5.08\pm0.84$ & 0.17 & 197 & 0.036 & $8.350\pm0.001$ & $1.176\pm0.014$ & $1.14\pm0.01$ & 8.950 & 0.870 & 0.300 & $6.431\pm0.030$ & $0.583\pm0.063$ & $0.130\pm0.062$ \\
47 & 56074& 2.20 & -1.12&10 & $2.87\pm0.91$ & 0.32 & 348 & 0.102 & $8.650\pm0.001$ & $0.955\pm0.011$ & $0.94\pm0.01$ & 9.170 & 0.780 & 0.260 & $6.837\pm0.020$ & $0.506\pm0.039$ & $0.145\pm0.047$ \\
48 & 56551& 2.20 & -0.59&11 & $3.45\pm0.99$ & 0.29 & 290 & 0.032 & $8.770\pm0.002$ & $1.077\pm0.023$ & $1.04\pm0.02$ &   $-$ &   $-$ &   $-$ & $7.048\pm0.023$ & $0.448\pm0.050$ & $0.447\pm0.047$ \\
49 & 57982& 2.60 & -1.04&10 & $1.10\pm1.19$ & 1.08 &520* & 0.071 & $9.580\pm0.002$ & $0.791\pm0.007$ & $0.82\pm0.01$ & 9.980 & 0.660 & 0.260 & $7.945\pm0.026$ & $0.447\pm0.043$ & $0.124\pm0.038$ \\
50 & 57983& 2.50 & -2.25& 1 & $2.02\pm1.10$ & 0.54 &522* & 0.021 & $9.590\pm0.002$ & $0.725\pm0.040$ & $0.78\pm0.03$ & 9.960 & 0.610 & 0.210 & $8.126\pm0.020$ & $0.409\pm0.070$ & $0.119\pm0.070$ \\
\hline
\end{tabular}
{
\\
}
}
\end{table*}
\end{landscape}

\setcounter{table}{0}
\begin{landscape} 
\begin{table*}
\setlength{\tabcolsep}{1.4pt}
\center
{\tiny
\begin{tabular}{crcccccccccccccccc}
\hline
ID & Hip & $\log g$ & $[M/H]$ & Ref. & $\pi$ & $\sigma_{\pi}/{\pi}$ & $d$ & $E_{d}(B-V)$ & $V$ & $B-V$ & $V-I$ & $g$ & $g-r$ & $r-i$ & $J$ & $J-H$ & $H-K_{s}$ \\
   &        &      & (dex)   &      & (mas) &  & (pc) & (mag) & (mag) & (mag) & (mag) & (mag) & (mag) & (mag) & (mag) & (mag) & (mag) \\
\hline  
51 & 58357& 2.50 & -0.82&25 & $6.85\pm0.99$ & 0.14 & 146 & 0.019 & $8.350\pm0.001$ & $0.860\pm0.015$ & $0.87\pm0.01$ & 8.830 & 0.680 & 0.220 & $6.681\pm0.019$ & $0.420\pm0.031$ & $0.102\pm0.030$ \\
52 & 59109& 3.00 & -2.62&21 & $4.87\pm1.51$ & 0.31 & 205 & 0.036 &$10.000\pm0.003$ & $0.413\pm0.027$ & $0.48\pm0.03$ &10.130 & 0.250 & 0.050 & $9.064\pm0.024$ & $0.266\pm0.060$ & $0.059\pm0.060$ \\
53 & 59239& 2.40 & -1.20&23 & $2.88\pm0.74$ & 0.26 & 347 & 0.020 & $8.610\pm0.002$ & $0.660\pm0.015$ & $0.72\pm0.01$ &   $-$ &   $-$ &   $-$ & $7.152\pm0.026$ & $0.396\pm0.035$ & $0.082\pm0.030$ \\
54 & 59334& 2.50 & -0.51&26 & $4.91\pm0.92$ & 0.19 & 204 & 0.033 & $8.370\pm0.002$ & $0.974\pm0.005$ & $0.96\pm0.01$ & 8.850 & 0.780 & 0.270 & $6.506\pm0.018$ & $0.543\pm0.027$ & $0.116\pm0.031$ \\
55 & 60719& 2.30 & -2.30&14 & $5.73\pm0.67$ & 0.12 & 175 & 0.013 & $8.030\pm0.001$ & $0.607\pm0.014$ & $0.68\pm0.01$ & 8.340 & 0.480 & 0.150 & $6.623\pm0.026$ & $0.390\pm0.040$ & $0.080\pm0.039$ \\
56 & 61175& 2.80 & -0.21&19 & $3.65\pm0.81$ & 0.22 & 274 & 0.042 & $8.090\pm0.001$ & $1.040\pm0.010$ & $1.01\pm0.01$ & 8.620 & 0.810 & 0.290 & $6.278\pm0.024$ & $0.505\pm0.052$ & $0.138\pm0.048$ \\
57 & 62235& 2.54 & -1.44& 9 & $0.52\pm1.08$ & 2.08 &425* & 0.016 & $9.140\pm0.002$ & $0.672\pm0.015$ & $0.73\pm0.01$ & 9.460 & 0.510 & 0.140 & $7.852\pm0.020$ & $0.309\pm0.066$ & $0.161\pm0.066$ \\
58 & 62747& 2.87 & -1.25&27 & $4.33\pm0.86$ & 0.20 & 231 & 0.040 & $7.970\pm0.002$ & $0.800\pm0.012$ & $0.83\pm0.01$ & 8.390 & 0.660 & 0.260 & $6.347\pm0.019$ & $0.449\pm0.039$ & $0.112\pm0.038$ \\
59 & 63955& 2.20 & -0.26&11 & $1.54\pm0.82$ & 0.53 &316* & 0.071 & $8.500\pm0.002$ & $1.183\pm0.019$ & $1.14\pm0.02$ & 9.120 & 0.980 & 0.350 & $6.631\pm0.019$ & $0.471\pm0.036$ & $0.175\pm0.036$ \\
60 & 64115& 2.69 & -0.35&28 & $4.71\pm1.03$ & 0.22 & 212 & 0.032 & $8.350\pm0.003$ & $0.963\pm0.003$ & $1.01\pm0.01$ & 8.880 & 0.780 & 0.270 & $6.541\pm0.018$ & $0.498\pm0.044$ & $0.156\pm0.045$ \\
61 & 65819& 2.70 & -0.86&10 & $4.38\pm1.06$ & 0.24 & 228 & 0.025 & $9.010\pm0.002$ & $0.839\pm0.001$ & $0.86\pm0.01$ & 9.510 & 0.680 & 0.220 & $7.386\pm0.018$ & $0.512\pm0.040$ & $0.072\pm0.041$ \\
62 & 68246& 2.94 & -0.66&29 & $8.62\pm0.91$ & 0.11 & 116 & 0.050 & $8.610\pm0.002$ & $0.424\pm0.004$ & $0.49\pm0.01$ &   $-$ &   $-$ &   $-$ & $7.695\pm0.026$ & $0.259\pm0.039$ & $0.005\pm0.044$ \\
63 & 69089& 2.60 &  0.04&30 &$-0.43\pm0.85$ &-1.98 &344* & 0.015 & $8.680\pm0.002$ & $1.062\pm0.020$ & $1.03\pm0.02$ & 9.180 & 0.730 & 0.270 & $6.952\pm0.020$ & $0.412\pm0.033$ & $0.161\pm0.031$ \\
64 & 70519& 2.16 & -1.78&27 & $3.82\pm0.77$ & 0.20 & 262 & 0.054 & $7.660\pm0.001$ & $0.791\pm0.003$ & $0.82\pm0.01$ & 8.080 & 0.680 & 0.220 & $5.975\pm0.019$ & $0.517\pm0.033$ & $0.120\pm0.031$ \\
65 & 70647& 3.00 & -2.20& 7 & $0.89\pm1.23$ & 1.38 &419* & 0.085 & $9.110\pm0.002$ & $0.774\pm0.021$ & $0.81\pm0.01$ & 9.530 & 0.850 & 0.310 & $7.258\pm0.020$ & $0.481\pm0.047$ & $0.109\pm0.047$ \\
66 & 71087& 2.20 & -1.58& 1 & $2.34\pm1.39$ & 0.59 &586* & 0.020 & $9.840\pm0.002$ & $0.818\pm0.015$ & $0.84\pm0.01$ &10.110 & 0.610 & 0.220 & $8.241\pm0.023$ & $0.404\pm0.029$ & $0.093\pm0.025$ \\
67 & 71458& 2.50 & -2.50&31 & $6.09\pm1.08$ & 0.18 & 164 & 0.046 & $8.020\pm0.002$ & $0.640\pm0.015$ & $0.70\pm0.01$ & 8.370 & 0.610 & 0.220 & $6.568\pm0.019$ & $0.417\pm0.028$ & $0.086\pm0.029$ \\
68 & 71761& 3.00 & -0.92& 7 & $5.42\pm1.49$ & 0.27 & 185 & 0.053 & $9.220\pm0.003$ & $0.606\pm0.019$ & $0.68\pm0.02$ & 9.510 & 0.460 & 0.130 & $7.981\pm0.030$ & $0.314\pm0.055$ & $0.095\pm0.052$ \\
69 & 72561& 3.00 & -1.61&32 & $5.00\pm2.23$ & 0.45 & 200 & 0.031 &$11.050\pm0.007$ & $0.413\pm0.028$ & $0.48\pm0.03$ &11.110 & 0.270 & 0.050 &$10.062\pm0.023$ & $0.220\pm0.034$ & $0.053\pm0.033$ \\
70 & 78378& 2.17 & -0.77& 9 & $6.35\pm0.78$ & 0.12 & 157 & 0.121 & $8.020\pm0.002$ & $0.972\pm0.013$ & $0.96\pm0.01$ & 8.520 & 0.780 & 0.270 & $6.139\pm0.023$ & $0.474\pm0.039$ & $0.152\pm0.036$ \\
71 & 80822& 3.00 & -0.90&23 & $3.50\pm1.08$ & 0.31 & 286 & 0.042 & $9.060\pm0.002$ & $0.672\pm0.020$ & $0.73\pm0.02$ & 9.420 & 0.560 & 0.190 & $7.619\pm0.019$ & $0.348\pm0.033$ & $0.086\pm0.034$ \\
72 & 82315& 2.81 & -0.63& 9 & $0.40\pm1.22$ & 3.05 &377* & 0.123 & $8.880\pm0.003$ & $0.852\pm0.031$ & $0.87\pm0.02$ & 9.280 & 0.730 & 0.270 & $7.101\pm0.023$ & $0.416\pm0.046$ & $0.170\pm0.044$ \\
73 & 85855& 2.92 & -2.35&29 & $3.62\pm1.05$ & 0.29 & 276 & 0.041 & $8.940\pm0.002$ & $0.607\pm0.003$ & $0.68\pm0.01$ & 9.240 & 0.610 & 0.220 & $7.420\pm0.019$ & $0.394\pm0.025$ & $0.074\pm0.025$ \\
74 & 88977& 3.00 & -1.00&33 & $4.56\pm0.84$ & 0.18 & 219 & 0.206 & $8.120\pm0.001$ & $0.875\pm0.022$ & $0.92\pm0.02$ & 8.680 & 0.980 & 0.380 & $5.998\pm0.021$ & $0.475\pm0.042$ & $0.179\pm0.044$ \\
75 & 90659& 3.00 &  0.67&34 & $8.95\pm0.71$ & 0.08 & 112 & 0.023 & $8.130\pm0.001$ & $0.803\pm0.005$ & $0.83\pm0.01$ & 8.530 & 0.650 & 0.220 & $6.478\pm0.019$ & $0.451\pm0.031$ & $0.074\pm0.030$ \\
76 & 96248& 2.64 & -1.85& 9 & $4.97\pm0.71$ & 0.14 & 201 & 0.052 & $7.590\pm0.001$ & $0.548\pm0.012$ & $0.62\pm0.01$ & 7.850 & 0.500 & 0.150 & $6.252\pm0.018$ & $0.339\pm0.032$ & $0.083\pm0.031$ \\
77 & 97747& 3.00 & -1.40&35 & $6.29\pm2.02$ & 0.32 & 159 & 0.036 &$10.760\pm0.005$ & $0.740\pm0.040$ & $0.79\pm0.03$ &11.130 & 0.650 & 0.190 & $9.161\pm0.023$ & $0.430\pm0.041$ & $0.074\pm0.041$ \\
78 & 98883& 3.00 & -1.00& 7 & $5.46\pm1.32$ & 0.24 & 183 & 0.051 & $9.510\pm0.002$ & $0.507\pm0.007$ & $0.58\pm0.01$ & 9.800 & 0.390 & 0.150 & $8.416\pm0.017$ & $0.273\pm0.032$ & $0.069\pm0.031$ \\
79 & 98974& 2.25 & -1.79& 7 & $4.78\pm0.98$ & 0.21 & 209 & 0.129 & $8.530\pm0.001$ & $0.753\pm0.019$ & $0.80\pm0.01$ & 8.930 & 0.640 & 0.210 & $6.970\pm0.032$ & $0.451\pm0.045$ & $0.078\pm0.035$ \\
80 &103337& 2.30 & -2.26&25 & $1.81\pm1.59$ & 0.88 &695* & 0.033 &$10.210\pm0.003$ & $0.893\pm0.015$ & $0.90\pm0.01$ &10.640 & 0.680 & 0.220 & $8.511\pm0.024$ & $0.489\pm0.033$ & $0.083\pm0.043$ \\
81 &104191& 2.50 & -3.16&36 & $2.54\pm1.15$ & 0.45 & 394 & 0.025 & $9.090\pm0.002$ & $0.582\pm0.027$ & $0.65\pm0.03$ &   $-$ &   $-$ &   $-$ & $7.648\pm0.018$ & $0.396\pm0.040$ & $0.098\pm0.041$ \\
82 &105993& 2.30 & -1.57& 6 & $3.23\pm1.71$ & 0.53 &682* & 0.054 &$10.170\pm0.004$ & $0.719\pm0.015$ & $0.77\pm0.01$ &10.430 & 0.510 & 0.140 & $8.800\pm0.035$ & $0.319\pm0.074$ & $0.118\pm0.069$ \\
83 &109390& 2.20 & -1.34&21 & $4.18\pm1.30$ & 0.31 & 239 & 0.031 & $9.550\pm0.004$ & $0.892\pm0.010$ & $0.90\pm0.01$ &10.010 & 0.780 & 0.260 & $7.839\pm0.020$ & $0.524\pm0.033$ & $0.132\pm0.033$ \\
84 &109501& 2.93 & -0.81&24 & $6.48\pm1.84$ & 0.28 & 154 & 0.033 & $8.630\pm0.001$ & $0.454\pm0.020$ & $0.52\pm0.02$ & 8.970 & 0.360 & 0.100 & $7.642\pm0.023$ & $0.198\pm0.033$ & $0.045\pm0.032$ \\
85 &110271& 2.20 & -1.03&10 & $0.76\pm1.47$ & 1.93 &421* & 0.018 & $9.120\pm0.003$ & $0.753\pm0.007$ & $0.80\pm0.01$ & 9.530 & 0.640 & 0.210 & $7.642\pm0.024$ & $0.436\pm0.052$ & $0.089\pm0.052$ \\
86 &111228& 2.85 & -1.40& 9 & $4.04\pm1.01$ & 0.25 & 248 & 0.011 & $8.520\pm0.001$ & $0.685\pm0.015$ & $0.74\pm0.01$ & 8.870 & 0.610 & 0.210 & $7.041\pm0.027$ & $0.492\pm0.038$ & $0.070\pm0.038$ \\
87 &112457& 2.02 & -1.35& 9 & $1.23\pm0.90$ & 0.73 &310* & 0.013 & $8.460\pm0.002$ & $0.776\pm0.017$ & $0.81\pm0.01$ & 8.880 & 0.660 & 0.250 & $6.850\pm0.019$ & $0.445\pm0.026$ & $0.125\pm0.029$ \\
88 &112821& 2.80 & -0.55&37 & $8.72\pm0.59$ & 0.07 & 115 & 0.029 & $7.460\pm0.001$ & $0.647\pm0.007$ & $0.71\pm0.01$ & 7.800 & 0.470 & 0.170 & $6.265\pm0.020$ & $0.231\pm0.029$ & $0.099\pm0.027$ \\
89 &114502& 2.40 & -1.86& 1 & $3.21\pm1.09$ & 0.34 & 312 & 0.032 & $8.940\pm0.002$ & $0.718\pm0.029$ & $0.77\pm0.02$ & 9.270 & 0.630 & 0.210 & $7.397\pm0.019$ & $0.406\pm0.060$ & $0.118\pm0.059$ \\
90 &116285& 2.49 & -1.14& 9 & $0.76\pm1.28$ & 1.68 &433* & 0.009 & $9.180\pm0.002$ & $0.689\pm0.004$ & $0.75\pm0.01$ & 9.500 & 0.610 & 0.220 & $7.736\pm0.024$ & $0.366\pm0.056$ & $0.123\pm0.054$ \\
91 &117168& 2.64 & -1.51& 9 & $2.48\pm0.74$ & 0.30 & 403 & 0.020 & $9.020\pm0.002$ & $0.571\pm0.011$ & $0.64\pm0.01$ & 9.310 & 0.450 & 0.160 & $7.747\pm0.023$ & $0.332\pm0.037$ & $0.077\pm0.039$ \\
\hline
\end{tabular}
{
\\
($^{*}$) The derived distance using the method explained in this paper\\
(1) \cite{Pilachowski1996}, (2) \cite{Smith1986}, (3) \cite{Burris2000}, (4) \cite{Spite1994}, (5) \cite{Kovacs1983}, (6) \cite{Gilroy1988}, (7) \cite{Gratton1983}, (8) \cite{Barbuy1989}, (9) \cite{Gratton2000}, (10) \cite{Ryan1995}, (11) \cite{Dominy1984}, (12) \cite{Ryan1998}, (13) \cite{Gratton1989}, (14) \cite{Luck1985}, (15) \cite{Luck1995}, (16) \cite{Fulbright1999}, (17) \cite{Kraft1992}, (18) \cite{Tomkin1999}, (19) \cite{Luck1991}, (20) \cite{Peterson1981}, (21) \cite{Fulbright2000}, (22) \cite{Drake1994}, (23) \cite{Gratton1984}, (24) \cite{Tomkin1992}, (25) \cite{Cavallo1997}, (26) \cite{Shetrone1996}, (27) \cite{Gratton1991}, (28) \cite{Clementini1999}, (29) \cite{Axer1994}, (30) \cite{LuckChallener1995}, (31) \cite{Spite1980}, (32) \cite{Spite1993}, (33) \cite{Leep1981}, (34) \cite{Mishenina1995}, (35) \cite{Carney1997}, (36) \cite{Ryan1991}, (37) \cite{Krishnaswamy1985}.
}
}
\end{table*}
\end{landscape}

We have omitted the indices $\infty$ and $d$ from the color excess $E(B-V)$ in the equations. 
However, we use the terms 'model' for the color excess of \cite{Schlegel98} and 
``reduced'' the color excess corresponding to distance $d$. The total absorption 
$A_{d}$ used in this section and the classical total absorption $A_{V}$ have the same meaning. 

The distances of 65 stars whose relative parallax errors are small were evaluated by using 
their original parallaxes. However, for one star with negative trigonometric parallax and 
for 25 stars with relative parallax errors $\sigma_{\pi}/\pi>0.5$ the following formula was used for this purpose: 

\begin{equation}
V-M_{V}-A_{d}(b)=5 \log d - 5.
\end{equation}
The absolute magnitudes for this sub--sample of stars were
adopted as $M_{V}=1$. The distances derived using this method are denoted with a star 
superscript in Table 1. As the total absorptions for the model and distance 
to the star are different, the distance to a star in this category (a total of 26 stars), 
as well as its total absorption, $A_{d}(b)$, and color excess, $E_{d}(B-V)$, could be 
evaluated by iterating Eqs. (1)--(4).

We de--reddened the colors and magnitudes by using the $E_{d}(B-V)$ color index of 
the stars evaluated using the procedures explained above and the equations of 
\cite{Yadav04, Fiorucci03, Fan1999} for $V-I$ color, and for 2MASS and SDSS 
photometries, respectively. The related equations are given in the following:

\begin{eqnarray}
V_{0}=V-3.1 E_{d}(B-V),\nonumber\\
(B-V)_{0}=(B-V) - E_{d}(B-V),\nonumber\\
(V-I)_{0}=(V-I) - 1.250E_{d}(B-V),\nonumber\\
J_{0}= J-0.887E_{d}(B-V),\nonumber\\
(J-H)_{0}= (J-H)-–0.322E_{d}(B-V),\nonumber\\
(H- K_{s})_{0}= (H-K_{s})–-0.183E_{d}(B-V),\nonumber\\
g_{0}=g-1.199 E_{d}(B-V),\nonumber\\
(g-r)_{0}=(g-r)–-0.341 E_{d}(B-V),\nonumber\\
(r-i)_{0}=(r-i)–-0.219 E_{d}(B-V).
\end{eqnarray}

The two color diagrams of the star sample for three photometric systems, i.e. 
$BVI$, $gri$ and $JHK_{s}$, compared to the calibrations of \cite{Pickles98}, 
\cite{Covey08} and \cite{Straizys09} are given in Fig. 3, respectively. The 
valid color index intervals for the transformations for giants are as follows: 
$0.25<(B-V)_{0}<1.35$, $0.25<(V - I)_{0}<1.35$, $0.10<(g - r)_{0}<0.95$, 
$0<(r - i)_{0}<0.35$, $0.05<(J - H)_{0}<0.60$, $0<(H - K_{s})_{0}<0.45$. 
The metallicity distribution of the sample covers all three populations, i.e. 
thin and thick disks, and halo (Fig. 4). The modes for the distributions of 
the whole sample (91 stars) and for the sample of stars (82 stars) which 
do not supply the relative parallax error condition ($\sigma_{\pi} / \pi<0.5$) are 
-1.23 and -1.21 dex, respectively. The corresponding medians of these 
distributions are close to the modes (-1.18 and -1.12 dex) indicating 
two Gaussian distributions.

\begin{figure}
\begin{center}
\includegraphics[scale=0.80, angle=0]{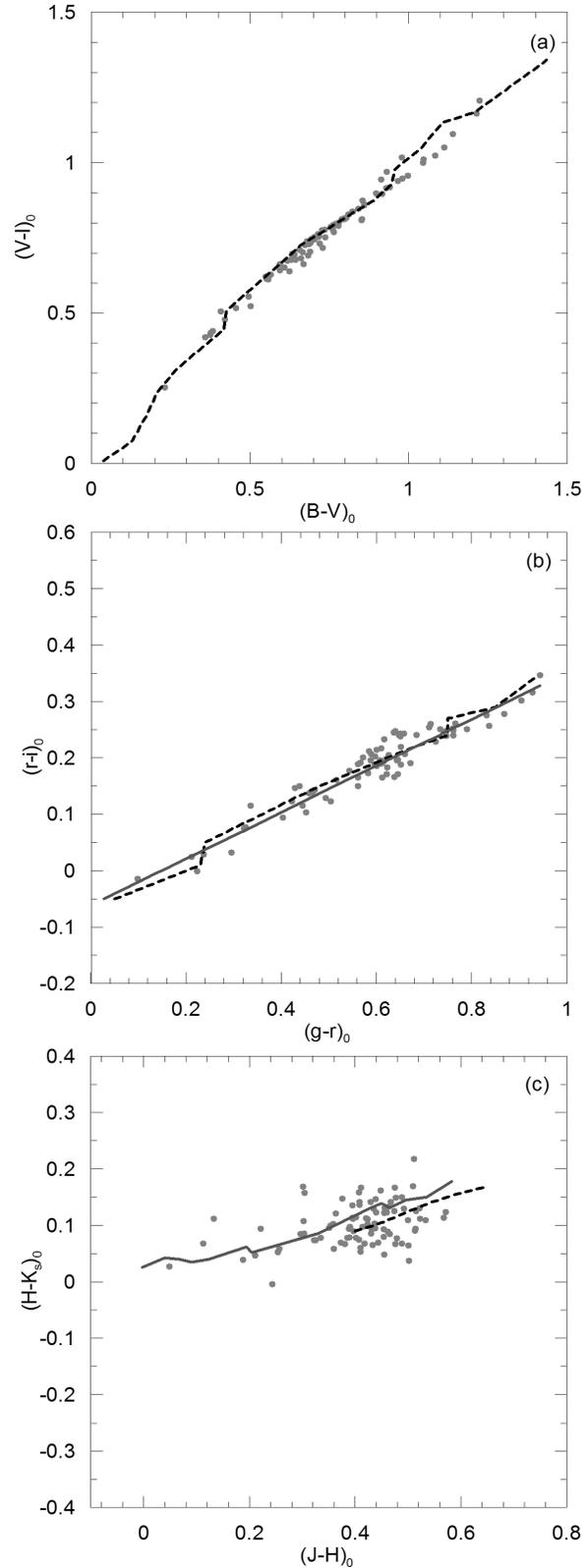}
\caption[] {Two--color diagrams of the sample stars. In all panels, the grey circles 
correspond to the positions of our sample stars. The dashed lines in panels (a), 
(b) and (c) are the calibrations of \cite{Pickles98}, \cite{Covey08}, and \cite{Straizys09}, 
respectively. (The continuous lines in panels (b) and (c) correspond to the two--color 
calibrations evaluated by the converted colors in Table 4. See section 4).}
\end{center}
\end{figure}

\begin{figure}
\begin{center}
\includegraphics[scale=0.40, angle=0]{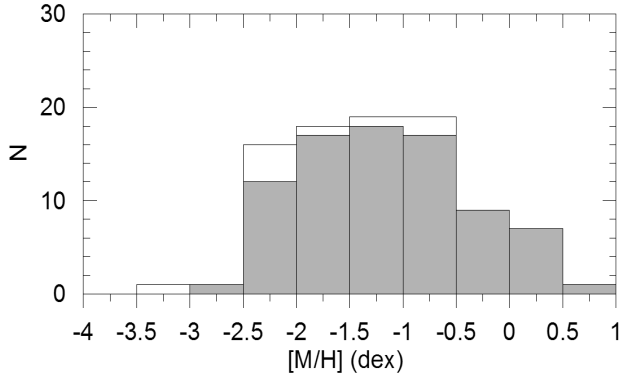}
\caption[] {Metalicity distributions for two samples: (1) for 91 late--type giant stars 
used in the transformations between $BVI$ and $JHK_{s}$ magnitudes (white area) and 
(2) for 82 giant stars used in the transformations between $gri$ and 
$JHK_{s}$ magnitudes (grey area).}
\end{center}
\end{figure}

\section{Results}
\subsection{Transformations between 2MASS and Johnson--Cousins photometry}

We used the following general equations and derived 12 sets of transformations 
between 2MASS and Johnson--Cousins $BVI$. Eqs. (6)--(11) transform 2MASS 
colors and magnitudes into $BVI$ magnitudes, whereas Eqs. (12)--(17) are their 
inverse transformations. The transformations are either metallicity dependent (Eqs. 
(6)--(8) and (12)--(14)) or independent of metallicity (Eqs. (9)--(11) and (15)--(17)). 
In this work, we followed a procedure different than the one in \cite{Bilir08}, used 
for the dwarfs. \cite{Bilir08} separated the sample of stars into metal--rich, 
intermediate metallicity and metal--poor sub--samples and obtained transformations 
for each sub--sample, whereas we adopted the metallicity as an additional term in 
Eqs. (6)--(8) and (12)--(14). This approach can be explained by the fact that 
stars change their positions in two color diagrams by shifting an amount 
proportional to their metallicities. The general equations are

\begin{equation}
(B-J)_{0}=a_{1}(J-H)_{0} + b_{1}(H-K_{s})_{0} + c_{1}[M/H] + d_{1},
\end{equation}
\begin{equation}
(V-J)_{0}=a_{2}(J-H)_{0} + b_{2}(H-K_{s})_{0} + c_{2}[M/H] + d_{2},
\end{equation}
\begin{equation}
(I-J)_{0}=a_{3}(J-H)_{0} + b_{3}(H-K_{s})_{0} + c_{3}[M/H] + d_{3},
\end{equation}
\begin{equation}
(B-J)_{0}=a_{4}(J-H)_{0} + b_{4}(H-K_{s})_{0} + d_{4},
\end{equation}
\begin{equation}
(V-J)_{0}=a_{5}(J-H)_{0} + b_{5}(H-K_{s})_{0} + d_{5},
\end{equation}
\begin{equation}
(I-J)_{0}=a_{6}(J-H)_{0} + b_{6}(H-K_{s})_{0} + d_{6},
\end{equation}
\begin{equation}
(V-J)_{0}=e_{1}(B-V)_{0} + f_{1}(V-I)_{0} + g_{1}[M/H] + h_{1},
\end{equation}
\begin{equation}
(V-H)_{0}=e_{2}(B-V)_{0} + f_{2}(V-I)_{0} + g_{2}[M/H] + h_{2},
\end{equation}
\begin{equation}
(V-K_{s})_{0}=e_{3}(B-V)_{0} + f_{3}(V-I)_{0} + g_{3}[M/H] + h_{3},
\end{equation}
\begin{equation}
(V-J)_{0}=e_{4}(B-V)_{0} + f_{4}(V-I)_{0} + h_{4},
\end{equation}
\begin{equation}
(V-H)_{0}=e_{5}(B-V)_{0} + f_{5}(V-I)_{0} + h_{5},
\end{equation}
\begin{equation}
(V-K_{s})_{0}=e_{6}(B-V)_{0} + f_{6}(V-I)_{0} + h_{6}.
\end{equation}
The numerical values of the coefficients in Eqs. (6)--(17) are given in Table 2. 

\begin{table}
\setlength{\tabcolsep}{1pt}
\center \caption{Coefficients for the transformation Eqs. (6)--(17) 
in column matrix. The subscript $i$ = 1, 2 and 3 or $i$ = 4, 5 and 6 correspond 
to the same number that denotes the equations. $R$ and $s$ are the correlation 
coefficient and standard deviation for  each category, respectively.}
\begin{tabular}{cccc}
\hline
\multicolumn{4}{c}{Equations dependent on metallicity, $i$ = 1, 2, 3}\\ 
                                                Coefficient &    $(B-J)_{0}$ & $(V-J)_{0}$ &  $(I-J)_{0}$ \\
\hline
                                              $a_{i}$  & $3.732\pm0.163$ & $2.328\pm0.102$ & $1.097\pm0.089$ \\
					     $b_{i}$   & $2.125\pm0.312$ & $1.128\pm0.195$ & $0.306\pm0.170$ \\
					     $c_{i}$   & $0.114\pm0.019$ & $0.033\pm0.012$ & $-0.031\pm0.011$ \\
					     $d_{i}$   & $0.592\pm0.071$ & $0.435\pm0.044$ & $0.185\pm0.038$ \\
					           $R$ &      0.950      &      0.945      &      0.832 \\
					           $s$ &      0.142      &      0.089      &      0.077 \\
\hline
\multicolumn{4}{c}{Equations independent from metallicity, $i$ = 4, 5, 6}\\
					       Coefficient &  $(B-J)_{0}$  &    $(V-J)_{0}$  &  $(I-J)_{0}$ \\
\hline
                                             $a_{i}$   & $3.671\pm0.191$ & $2.310\pm0.105$ & $1.114\pm0.092$ \\
					     $b_{i}$   & $2.518\pm0.359$ & $1.242\pm0.198$ & $0.198\pm0.173$ \\
					     $d_{i}$   & $0.436\pm0.077$ & $0.390\pm0.042$ & $0.228\pm0.037$ \\
					           $R$ &      0.929      &      0.940      &      0.812 \\
					           $s$ &      0.167      &      0.092 	    &     0.080 \\
\hline
\multicolumn{4}{c}{Equations dependent on metallicity, $i$ = 1, 2, 3}\\ 
					      Coefficient &  $(V-J)_{0}$   &  $(V-H)_{0}$  & $(V-K_{s})_{0}$ \\
\hline
                                              $e_{i}$   & $1.080\pm0.433$ & $1.454\pm0.590$ & $1.791\pm0.592$ \\
					      $f_{i}$   & $0.379\pm0.495$ & $0.504\pm0.674$ & $0.294\pm0.676$ \\
					      $g_{i}$   & $-0.082\pm0.015$ & $-0.127\pm0.020$ & $-0.129\pm0.020$ \\
					      $h_{i}$   & $0.279\pm0.078$ & $0.261\pm0.106$ & $0.276\pm0.106$ \\
					            $R$ &      0.933      &      0.931      &      0.940 \\
					            $s$ &      0.097      &      0.133      &      0.133 \\
\hline
\multicolumn{4}{c}{Equations independent from metallicity, $i$ = 4, 5, 6}\\
					      Coefficient &  $(V-J)_{0}$   &  $(V-H)_{0}$ & $(V-K_{s})_{0}$ \\
\hline
                                             $e_{i}$   & $0.397\pm0.484$ & $0.402\pm0.684$ & $0.716\pm0.690$ \\
					     $f_{i}$   & $1.018\pm0.560$ & $1.488\pm0.792$ & $1.299\pm0.799$ \\
					     $h_{i}$   & $0.396\pm0.087$ & $0.441\pm0.123$ & $0.460\pm0.124$ \\
					           $R$ &      0.907      &      0.896      &      0.908 \\
					           $s$ &      0.113      &      0.160      &      0.162 \\
\hline
\end{tabular} 
\end{table}

\subsection{Transformations between 2MASS and SDSS}                        

The transformations between 2MASS and SDSS, given in the following, 
have similar general equations:

\begin{equation}
(g-J)_{0} =k_{1}(J-H)_{0} + l_{1}(H-K_{s})_{0} + m_{1}[M/H] + n_{1}.
\end{equation}
\begin{equation} 
(r-J)_{0}=k_{2}(J-H)_{0} + l_{2}(H-K_{s})_{0} + m_{2}[M/H] + n_{2}.
\end{equation}
\begin{equation}
(i-J)_{0}=k_{3}(J-H)_{0}+ l_{3}(H-K_{s})_{0} + m_{3}[M/H] + n_{3}.
\end{equation}
\begin{equation}
(g-J)_{0}=k_{4}(J-H)_{0}+ l_{4}(H-K_{s})_{0} + n_{4}.
\end{equation}
\begin{equation}
(r-J)_{0}=k_{5}(J-H)_{0}+ l_{5}(H-K_{s})_{0} + n_{5}.
\end{equation}
\begin{equation}
(i-J)_{0}=k_{6}(J-H)_{0}+ l_{6}(H-K_{s})_{0} + n_{6}.
\end{equation}
\begin{equation}
(g-J)_{0}=o_{1}(g-r)_{0}+ p_{1}(r-i)_{0} + r_{1}[M/H] + s_{1}.
\end{equation}
\begin{equation}
(g-H)_{0}=o_{2}(g-r)_{0}+ p_{2}(r-i)_{0} + r_{2}[M/H] + s_{2}.
\end{equation}
\begin{equation}
(g-K_{s})_{0}=o_{3}(g-r)_{0}+ p_{3}(r-i)_{0} + r_{3}[M/H] + s_{3}.
\end{equation}
\begin{equation}
(g-J)_{0}=o_{4}(g-r)_{0}+p_{4}(r-i)_{0} + s_{4}.
\end{equation}
\begin{equation}
(g-H)_{0}=o_{5}(g-r)_{0}+ p_{5}(r-i)_{0} + s_{5}.
\end{equation}
\begin{equation}
(g-K_{s})_{0}=o_{6}(g-r)_{0}+ p_{6}(r-i)_{0} + s_{6}.
\end{equation}

One can see that, the equations which convert 2MASS colors and magnitudes 
into $gri$ magnitudes and their inverse transformations are either metallicity dependent 
or independent of metallicity. The numerical values of the coefficients in Eqs. 
(18)--(29) are given in Table 3. 

\begin{table}
\setlength{\tabcolsep}{1pt}
\center \caption{Coefficients for the transformation Eqs. (18)--(29) 
in column matrix. The subscript $i$ = 1, 2 and 3 or $i$ = 4, 5 and 6 correspond 
to the same number that denotes the equations. $R$ and $s$ are the correlation 
coefficient and standard deviation for  each category, respectively.}
\begin{tabular}{cccc}
\hline
\multicolumn{4}{c}{Equations dependent on metallicity, $i$ = 1, 2, 3}\\ 
					   Coefficient  & $(g-J)_{0}$     & $(r-J)_{0}$   &    $(i-J)_{0}$ \\
\hline
					      $k_{i}$   & $2.992\pm0.127$ & $1.743\pm0.087$ & $1.233\pm0.083$ \\
					     $l_{i}$   & $2.478\pm0.357$ & $1.250\pm0.245$ & $0.727\pm0.232$ \\
					     $m_{i}$   & $0.089\pm0.016$ & $0.054\pm0.011$ & $0.042\pm0.010$ \\
					     $n_{i}$   & $0.461\pm0.057$ & $0.462\pm0.039$ & $0.524\pm0.037$ \\
					           $R$ &       0.960     &      0.943      &      0.901 \\
		                                   $s$ &       0.106     &      0.073      &      0.069 \\
\hline
\multicolumn{4}{c}{Equations independent from metallicity, $i$ = 4, 5, 6}\\ 			
					               & $(g-J)_{0}$    & $(r-J)_{0}$   &    $(i-J)_{0}$ \\
\hline
					     $k_{i}$   & $2.923\pm0.148$ & $1.701\pm0.099$ & $1.200\pm0.090$ \\
					     $l_{i}$   & $3.031\pm0.400$ & $1.585\pm0.266$ & $0.991\pm0.243$ \\
					     $n_{i}$   & $0.329\pm0.061$ & $0.383\pm0.041$ & $0.461\pm0.037$ \\
					           $R$ &      0.943      &       0.925     &      0.879 \\
					           $s$ &      0.124      &       0.083     &      0.075 \\
\hline
\multicolumn{4}{c}{Equations dependent on metallicity, $i$ = 1, 2, 3}\\ 
					               &    $(g-J)_{0}$  &  $(g-H)_{0}$  & $(g-K_{s})_{0}$ \\
\hline
					     $o_{i}$   & $2.150\pm0.199$ & $2.767\pm0.270$ & $2.885\pm0.267$ \\
					     $p_{i}$   & $0.194\pm0.462$ & $0.097\pm0.627$ & $0.130\pm0.618$ \\
					     $r_{i}$   & $0.008\pm0.012$ & $-0.021\pm0.017$& $-0.015\pm0.016$ \\
					     $s_{i}$   & $0.531\pm0.050$ & $0.556\pm0.068$ & $0.590\pm0.067$ \\
					           $R$ &      0.977      &      0.972      &       0.975 \\
					           $s$ &      0.081      &      0.110      &      0.108 \\
\hline
\multicolumn{4}{c}{Equations independent from metallicity, $i$ = 4, 5, 6}\\ 
					               &    $(g-J)_{0}$  & $(g-H)_{0}$ & $(g-K_{s})_{0}$ \\
\hline
					     $o_{i}$   & $2.172\pm0.196$ & $2.706\pm0.267$ & $2.842\pm0.262$ \\
					     $p_{i}$   & $0.160\pm0.457$ & $0.185\pm0.626$ & $0.192\pm0.614$ \\
					     $s_{i}$   & $0.514\pm0.043$ & $0.600\pm0.058$ & $0.622\pm0.057$ \\
					           $R$ &      0.976      &      0.972      &      0.975 \\
					           $s$ &      0.081      &      0.110      &      0.108 \\
\hline
\end{tabular} 
\end{table}

\subsection{Residuals}

We compared the observed colors with those evaluated via Eqs. (6)--(29). 
The residuals are plotted versus observed $(B-V)_{0}$, $(J-H)_{0}$ or $(g-r)_{0}$ colors in Fig. 5. 
There are some small systematic deviations in the the red and blue ends of the $(B-V)_{0}$ color. 
However, they are negligible when compared to the accuracy of the transformations. Actually, 
the mean of the residuals are smaller than a thousandth and the standard deviations are close 
to 0.1 mag for all colors (Table 4). The ranges of the residuals for different colors are not 
the same. The residuals are larger for colors over wider wavelength intervals but they are smaller 
for metal dependent transformations than the corresponding ones for metal independent transformations. 
One can confirm this argument by comparing the ranges of the residuals in Fig. 5 and the standard 
deviations in Table 4.

\section{Discussion}

We present color transformations for late--type giants from 2MASS photometric system 
into the Johnson--Cousins\rq system and furthermore into the SDSS system. 
Thus, this is the complement of the paper of \cite{Bilir08} where 
transformations were carried out for dwarfs. We adopted the following steps 
in order to obtain accurate transformations: (1) the sample was selected 
by means of the star surface gravities, i.e. $2 < \log g \leq 3$, (2) the photometric data was 
de--reddened, (3) stars in the sample were selected according to their photometric data quality,  
(4) transformations were constructed as two--color dependent equations. Step (3) is especially important 
for 2MASS data because the inherent errors are larger relative to those in other 
photometries. Additionally, the transformations are given either as a function of 
metallicity or independent from metallicity. However, in the case of metallicity 
dependent equations the procedure is different than the one of Bilir et al.\rq s (2008). 
Instead of separating the star sample into different metallicity intervals, we added 
the metallicity as a term into the transformation equations. 

\begin{figure*}
\begin{center}
\includegraphics[scale=0.80, angle=0]{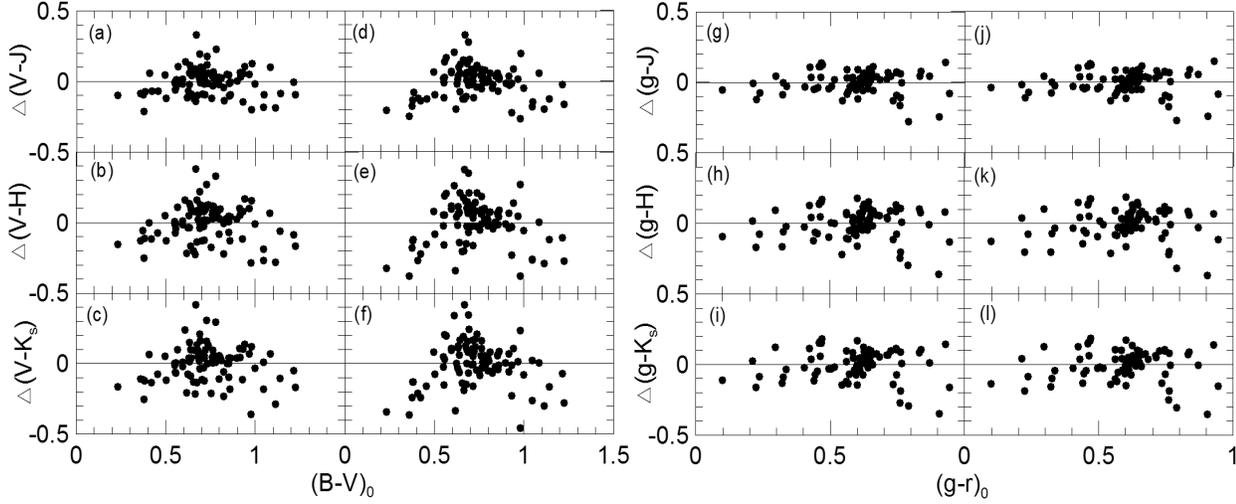}
\caption[] {Color residuals for two sets of transformations. The notation used is 
$\Delta$(color)=(evaluated color)-–(measured color). The first and third columns 
correspond to the metal dependent transformations, whereas residuals for metallicity 
independent transformations are given in the second and fourth columns.}
\end{center}
\end{figure*}

\begin{table*}
\center \caption{Averages and standard deviations ($s$) for the residuals 
for different colors in transformation Eqs. (6)--(29). The notation 
used is $\Delta$(color)=(measured color)--(evaluated color).}
\begin{tabular}{rcccccc}
\hline
\multicolumn{7}{c}{For metallicity dependent transformation equations}            \\
\hline
           & $\Delta(B-J)$ & $\Delta(V-J)$ & $\Delta(I-J)$ & $\Delta(g-J)$ & $\Delta(r-J)$ & $\Delta(i-J)$ \\
\hline
   Average &     0.0007 &     0.0001 &     0.0006 &     0.0001 &     0.0010 &    -0.0004 \\
       $s$ &      0.142 &      0.089 &      0.077 &      0.106 &      0.073 &      0.069 \\
\hline
           & $\Delta(V-J)$ & $\Delta(V-H)$ & $\Delta(V-K_{s})$ & $\Delta(g-J)$ & $\Delta(g-H)$ & $\Delta(g-K_{s})$ \\
\hline
   Average &     0.0001 &    -0.0004 &     0.0002 &    -0.0007 &     0.0001 &     0.0002 \\
       $s$ &      0.097 &      0.133 &      0.133 &      0.081 &      0.110 &      0.108 \\
\hline
\multicolumn{7}{c}{For transformation equations independent from metallicity}  \\
\hline
           & $\Delta(B-J)$ & $\Delta(V-J)$ & $\Delta(I-J)$ & $\Delta(g-J)$ & $\Delta(r-J)$ & $\Delta(i-J)$ \\
\hline
   Average &     0.0003 &     0.0001 &     0.0001 &     0.0003 &     0.0001 &     0.0003 \\
       $s$ &      0.167 &      0.092 &      0.080 &      0.124 &      0.083 &      0.075 \\
\hline
           & $\Delta(V-J)$ & $\Delta(V-H)$ & $\Delta(V-K_{s})$ & $\Delta(g-J)$ & $\Delta(g-H)$ & $\Delta(g-K_{s})$ \\
\hline
   Average &    -0.0005 &    -0.0001 &    -0.0001 &     0.0004 &    -0.0001 &    -0.0005 \\
       $s$ &      0.113 &      0.160 &      0.162 &      0.081 &      0.110 &      0.108 \\
\hline
\end{tabular}  

\end{table*}

The coefficients of the color terms in the same transformation equation are significant
(see Tables 2 and 3) which indicate that the transformations are two--color 
dependent. On the other hand, the correlation coefficients for the metallicity dependent 
transformations are larger than the ones of metallicity independent transformations indicating that 
the transformations actually vary with metallicity. The mean of the residuals, i.e. the 
color differences between those measured and those calculated, are smaller than 0.001 mag 
(except for the color $(r-J)$ which is equal to 0.001 mag). Additionally, the deviations 
of the measured colors from the calculated ones are absolutely smaller ($s\leq 0.16$) than 
the ones cited by \cite{Bilir08} for dwarfs ($0.2 < s\leq 0.3$).

We applied our transformation equations to the synthetic $BVI$ data of \cite{Pickles98} 
which cover stars of a wide spectral type range, and obtained $J-H$, 
$H-K_{s}$, $g-r$ and $r-i$ colors (Table 5) which gave us the chance to compare 
the two color diagrams in different studies. In all panels of Fig. 3, the grey circles 
correspond to the positions of our sample stars, whereas the dashed lines in panels (a), (b) 
and (c) represent the calibrations of \cite{Pickles98}, \cite{Covey08} and \cite{Straizys09}, 
respectively. Finally, the continuous lines in panels (b) and (c) correspond to the 
two--color calibrations evaluated using the converted colors in Table 5. 
The deviation between the dashed and continuous lines in panel (b) originates from 
deviations of the positions of our sample stars from the synthetic $(B-V)$ and $(V-I)$ colors 
in panel (a). But the difference between two lines in panel (c) may be due to the uncertainty of the 
calibration itself, in addition to the mentioned deviations. However, we can argue that the 
transformations derived for late type giants in this work could be applied with sufficient accuracy. 

\begin{table*}
\center \caption{Synthetic data taken from Pickles\rq (1998) (columns 1-–5) and 
$J-H$, $H-K_{s}$, $g-r$ and $r-i$ colors evaluated by the 
corresponding transformations derived in this work.}
\begin{tabular}{ccccccccc}
\hline
Spectral Type & $[M/H]$ (dex) & $T_{eff}$ (K) & $B-V$ & $V-I_{c}$ & $J-H$ & $H-K_{s}$ & $g-r$ & $r-i$ \\
\hline
     A0III &          0 &       9572 &      0.037 &      0.008 &     -0.003 &      0.026 &      0.027 &     -0.050 \\
     A3III &          0 &       8974 &      0.130 &      0.076 &      0.040 &      0.043 &      0.102 &     -0.019 \\
     A5III &          0 &       8453 &      0.175 &      0.164 &      0.068 &      0.040 &      0.133 &     -0.006 \\
     A7III &          0 &       8054 &      0.210 &      0.240 &      0.091 &      0.035 &      0.156 &      0.003 \\
     F0III &          0 &       7586 &      0.271 &      0.317 &      0.123 &      0.040 &      0.202 &      0.022 \\
     F2III &          0 &       6839 &      0.417 &      0.445 &      0.194 &      0.062 &      0.317 &      0.069 \\
     F5III &          0 &       6531 &      0.425 &      0.507 &      0.204 &      0.052 &      0.318 &      0.069 \\
     G0III &      -0.22 &       5610 &      0.661 &      0.726 &      0.330 &      0.086 &      0.509 &      0.149 \\
     G5III &      -0.12 &       5164 &      0.886 &      0.872 &      0.428 &      0.131 &      0.690 &      0.223 \\
     G8III &       0.06 &       5012 &      0.946 &      0.927 &      0.449 &      0.139 &      0.733 &      0.240\\
     K0III &      -0.08 &       4853 &      0.957 &      0.978 &      0.466 &      0.132 &      0.740 &      0.244 \\
     K1III &       0.09 &       4656 &      1.036 &      1.044 &      0.496 &      0.145 &      0.800 &      0.268 \\
     K2III &       0.05 &       4457 &      1.105 &      1.132 &      0.535 &      0.150 &      0.853 &      0.290 \\
     K3III &      -0.02 &       4365 &      1.210 &      1.169 &      0.582 &      0.177 &      0.943 &      0.327 \\
\hline
\end{tabular}  

\end{table*}       

\section{Acknowledgments}
We thank the anonymous referee for a thorough report and useful comments that helped in improving an early 
version of the paper. S. Karaali is grateful to the Beykent University for financial support. 
This publication makes use of data products from the Two Micron All
Sky Survey, which is a joint project of the University of
Massachusetts and the Infrared Processing and Analysis
Center/California Institute of Technology, funded by the National
Aeronautics and Space Administration and the National Science
Foundation.

This research has made use of the SIMBAD, NASA\rq s Astrophysics Data
System Bibliographic Services and the NASA/IPAC Extragalactic
Database (NED) which is operated by the Jet Propulsion Laboratory,
California Institute of Technology, under contract with the National
Aeronautics and Space Administration.


\begin{thebibliography}{99}
\bibitem[Abazajian et al.(2004)]{Abazajian04} Abazajian K., Adelman-McCarthy J.K., Agüeros M.A., Allam S.S., Anderson K., Anderson S.F., Annis J., Bahcall N.A., Baldry I.K., Bastian S., and 143 coauthors: 2004, AJ 128, 502

\bibitem[Axer et al.(1994)]{Axer1994} Axer M., Fuhrmann K., Gehren T.: 1994, A\&A 291, 895

\bibitem[Bahcall \& Soneira(1980)]{Bahcall80} Bahcall J.N., Soneira R.M.: 1980, ApJS 44, 73

\bibitem[Barbuy \& Erdelyi-Mendes(1989)]{Barbuy1989} Barbuy B., Erdelyi-Mendes M.: 1989, A\&A 214, 239

\bibitem[Bilir et al.(2008)]{Bilir08} Bilir S., Ak S., Karaali S., Cabrera-Lavers A., Chonis T.S., Gaskell C.M.: 2008, MNRAS 384, 1178

\bibitem[Burris et al.(2000)]{Burris2000} Burris D.L., Pilachowski C.A., Armandroff T.E., Sneden C., Cowan J.J., Roe H.: 2000, ApJ 544, 302

\bibitem[Cabrera-Lavers et al.(2008)]{Cabrera2008} Cabrera-Lavers A., Gonz{\'a}lez-Fern{\'a}ndez C., Garz{\'o}n F., Hammersley P.~L., \& L{\'o}pez-Corredoira M.: 2008, A\&A 491, 781

\bibitem[Carney et al.(1997)]{Carney1997} Carney B.W., Wright J.S., Sneden C., Laird J.B., Aguilar L.A., Latham D.W.: 1997, AJ 114, 363

\bibitem[Cavallo et al.(1997)]{Cavallo1997} Cavallo R.M., Pilachowski C.A., Rebolo R.: 1997, PASP 109, 226

\bibitem[Cayrel de Strobel et al.(2001)]{Cayrel01} Cayrel de Strobel G., Soubiran C., Ralite N.: 2001, A\&A 373, 159

\bibitem[Chonis \& Gaskell(2008)]{Chonis08} Chonis T.S., Gaskell C.M.: 2008, AJ 135, 264

\bibitem[Clementini et al.(1999)]{Clementini1999} Clementini G., Gratton R.G., Carretta E., Sneden C.: 1999, MNRAS 302, 22

\bibitem[Covey et al.(2008)]{Covey08} Covey K.R., Hawley S.L., Bochanski J.J., West A.A., Reid I.N., Golimowski D.A., Davenport J.R.A., Henry T., Uomoto A., Holtzman J.A.: 2008, AJ 136, 1778

\bibitem[Cutri et al.(2003)]{Cu03} Cutri R. M., Skrutskie M.F., Van Dyk S., Beichman C.A., Carpenter J.M., Chester T., Cambresy L., Evans T., Fowler J., Gizis J., Howard E., Huchra J., Jarrett T., Kopan E.L., Kirkpatrick J.D., Light R.M, Marsh K.A., McCallon H., Schneider S., Stiening R., Sykes M., Weinberg M., Wheaton W.A., Wheelock S., Zacarias N.: 2003, 2MASS All-Sky Catalog of Point Sources, CDS/ADC Electronic Catalogues 2246

\bibitem[Davenport et al.(2007)]{Davenport07} Davenport J.R.A., Bochanski J.J., Covey K.R., Hawley S.L.,West A.A., Schneider D.P.: 2007, AJ 134, 2430
 
\bibitem[Dominy(1984)]{Dominy1984} Dominy J.F.: 1984, ApJS 55, 27

\bibitem[Drake \& Lambert(1994)]{Drake1994} Drake J.J., Lambert D.L.: 1994, ApJ 435, 797

\bibitem[ESA(1997)]{ESA97}ESA: 1997, The Hipparcos and Tycho Catalogues ESA SP-1200. ESA, Noordwijk  

\bibitem[Fan(1999)]{Fan1999} Fan X.: 1999, AJ 117, 2528

\bibitem[Fiorucci \& Munari(2003)]{Fiorucci03} Fiorucci M., Munari U.: 2003, A\&A 401, 781

\bibitem[Fukugita et al.(1996)]{Fukugita96} Fukugita M., Ichikawa T., Gunn J.E., Doi M., Shimasaku K., Schneider D.P.: 1996, AJ 111, 1748

\bibitem[Fulbright \& Kraft(1999)]{Fulbright1999} Fulbright J.P., Kraft R.P.: 1999, AJ 118, 527

\bibitem[Fulbright(2000)]{Fulbright2000} Fulbright J.P.: 2000, AJ 120, 1841

\bibitem[Gilroy et al.(1988)]{Gilroy1988} Gilroy K.K., Sneden C., Pilachowski C.A., Cowan J.J.: 1988, ApJ 327, 298

\bibitem[Gratton(1983)]{Gratton1983} Gratton R.G.: 1983, A\&A 123, 289

\bibitem[Gratton \& Ortolani(1984)]{Gratton1984} Gratton R.G., Ortolani S.: 1984, A\&A 137, 6

\bibitem[Gratton(1989)]{Gratton1989} Gratton R.G.: 1989, A\&A 208, 171

\bibitem[Gratton \& Sneden(1991)]{Gratton1991} Gratton R.G., Sneden C.: 1991, A\&A 241, 501

\bibitem[Gratton et al.(2000)]{Gratton2000} Gratton R.G., Sneden C., Carretta E., Bragaglia A.: 2000, A\&A 354, 169

\bibitem[Gunn et al.(1998)]{Gunn98} Gunn J.E., Carr M., Rockosi C., Sekiguchi M., Berry K., Elms B., de Haas E., Ivezic Z., Knapp G., Lupton R., and 30 coauthors: 1998, AJ 116, 3040

\bibitem[Hogg et al.(2001)]{Hogg01} Hogg D.W., Finkbeiner D.P., Schlegel D.J., Gunn J.E.: 2001, AJ 122, 2129

\bibitem[Jordi et al.(2006)]{Jordi06} Jordi K., Grebel E.K., Ammon K.: 2006, A\&A 460, 339

\bibitem[Karaali et al.(2005)]{Karaali05} Karaali S., Bilir S., Tun\c{c}el S.: 2005, PASA 22, 24

\bibitem[Kovacs(1983)]{Kovacs1983} Kovacs N.: 1983, A\&A 120, 21

\bibitem[Kraft et al.(1992)]{Kraft1992} Kraft R.P., Sneden C., Langer G.E., Prosser C.F.: 1992, AJ 104, 645

\bibitem[Krishnaswamy \& Sneden(1985)]{Krishnaswamy1985} Krishnaswamy K., Sneden C.: 1985, PASP 97, 407

\bibitem[Landolt(1992)]{Landolt92} Landolt A.U.: 1992, AJ 104, 340

\bibitem[Leep \& Wallerstein(1981)]{Leep1981} Leep E.M., Wallerstein G.: 1981, MNRAS 196, 543

\bibitem[L{\'o}pez-Corredoira et al.(2002)]{Lopez2002} L{\'o}pez-Corredoira M., Cabrera-Lavers A., Garz{\'o}n F., \& Hammersley P.~L.: 2002, A\&A 394, 883 

\bibitem[Lucas et al.(2008)]{Lucas2008} Lucas P.W., Hoare M.G., Longmore A., Schröder A.C., Davis C.J., Adamson A., Bandyopadhyay R.M., de Grijs R., Smith M., Gosling A., and 21 coauthors: 2008, MNRAS 391, 136 

\bibitem[Luck \& Bond(1985)]{Luck1985} Luck R.E., Bond H.E.: 1985, ApJ 292, 559

\bibitem[Luck \& Bond(1991)]{Luck1991} Luck R.E., Bond H.E.: 1991, ApJS 77, 515

\bibitem[Luck \& Wepfer(1995)]{Luck1995} Luck R.E., Wepfer G.G.: 1995, AJ 110, 2425

\bibitem[Luck \& Challener(1995)]{LuckChallener1995} Luck R.E., Challener S.L.: 1995, AJ 110, 2968

\bibitem[Lupton et al.(2001)]{Lupton01} Lupton R.H., Gunn J.E., Ivezic Z., Knapp G.R., Kent S., Yasuda N.: 2001, in ASP Conf. Ser.: Astronomical Data Analysis Software and Systems X, ed. F. R. Harden Jr., F.A. Primini and H.E. Payne, 238, 269

\bibitem[Marshall et al.(2006)]{Marshall06} Marshall D.J., Robin A.C., Reyl\'{e} C., Schultheis M., Picaud S.: 2006, A\&A 453, 635

\bibitem[Mishenina et al.(1995)]{Mishenina1995} Mishenina T.V., Klochkova V.G., Panchuk V.E.: 1995, A\&AS 109, 471 

\bibitem[Ofek(2008)]{Ofek08} Ofek E.O.: 2008, PASP 120, 1128

\bibitem[Peterson(1981)]{Peterson1981} Peterson R.C.: 1981, ApJS 45, 421 

\bibitem[Pickles(1998)]{Pickles98} Pickles A.J.: 1998, PASP 110, 863

\bibitem[Pilachowski et al.(1996)]{Pilachowski1996} Pilachowski C.A., Sneden C., Kraft, Robert P.: 1996, AJ 111, 1689

\bibitem[Rodgers et al.(2006)]{Rodgers06} Rodgers C.T., Canterna R., Smith J.A., Pierce M.J., Tucker D.L.: 2006, AJ 132, 989

\bibitem[Ryan et al.(1991)]{Ryan1991} Ryan S.G., Norris J.E., Bessell M.S.: 1991, AJ 102, 303

\bibitem[Ryan \& Lambert(1995)]{Ryan1995} Ryan S.G., Lambert D.L.: 1995, AJ 109, 2068

\bibitem[Ryan \& Deliyannis(1998)]{Ryan1998} Ryan S.G., Deliyannis C.P.: 1998, ApJ 500, 398 

\bibitem[Schlegel et al.(1998)]{Schlegel98} Schlegel D.J., Finkbeiner D.P., Davis M.: 1998, ApJ 500, 525

\bibitem[Shetrone(1996)]{Shetrone1996} Shetrone M.D.: 1996, AJ 112, 1517

\bibitem[Skrutskie et al.(2006)]{2mass06} Skrutskie M.F., Cutri R.M., Stiening R., Weinberg M.D., Schneider S., Carpenter J.M., Beichman C., Capps R., Chester T., Elias J., and 21 coauthors et al.: 2006, AJ 131, 1163

\bibitem[Smith \& Lambert(1986)]{Smith1986} Smith V.V., Lambert D.L.: 1986, ApJ 303, 226

\bibitem[Smith et al.(2002)]{Smith02} Smith J.A., Tucker D.L., Kent S., Richmond M.W., Fukugita M., Ichikawa T., Ichikawa S., Jorgensen A.M., Uomoto A., Gunn J.E.: 2002, AJ 123, 2121

\bibitem[Smith et al.(2007)]{Smith07} Smith J.A., Tucker D.L., Allam S.S., Ivezic Z., Yanny B., Gunn J.E., Knapp G.R., Eisenstein D., Finkbeiner D., Fukugita M.: 2007, The Future of Photometric, Spectrophotometric and Polarimetric Standardization, ASP Conference Series, Vol. 364, Edited by C. Sterken, San Francisco: Astronomical Society of the Pacific, p. 91 

\bibitem[Spite \& Spite(1980)]{Spite1980} Spite M., Spite F.: 1980, A\&A 89, 118

\bibitem[Spite et al.(1993)]{Spite1993} Spite M., Molaro P., Francois P., Spite F.: 1993, A\&A 271L, 1

\bibitem[Spite et al.(1994)]{Spite1994} Spite, M., Pasquini, L., Spite, F.: 1994, A\&A 290, 217

\bibitem[Straizys \& Lazauskaite(2009)] {Straizys09}Straizys V., Lazauskaite R.: 2009, BaltA 18, 19

\bibitem[Tomkin et al.(1992)]{Tomkin1992} Tomkin J., Lemke M., Lambert D.L., Sneden C.: 1992, AJ 104, 1568

\bibitem[Tomkin \& Lambert(1999)]{Tomkin1999} Tomkin J., Lambert D.L.: 1999, ApJ 523, 234

\bibitem[Tucker et al.(2006)]{Tucker06} Tucker D.L., Kent, S., Richmond, M.W., Annis, J., Smith, J.A., Allam, S.S., Rodgers, C.T., Stute, J.L., Adelman-McCarthy, J.K., Brinkmann, J., and 24 coauthors: 2006, AN 327, 821

\bibitem[van Leeuwen(2007)]{vanLeeuwen2007} van Leeuwen F.: 2007, A\&A 474, 653

\bibitem[Walkowicz et al.(2004)]{Walkowicz04} Walkowicz L.M., Hawley S.L., West A.A.: 2004, PASP 116, 1105

\bibitem[West et al.(2005)]{West05} West A.A., Walkowicz L.M., Hawley S.L.: 2005, PASP 117, 706

\bibitem[Yadav \& Sagar(2004)]{Yadav04} Yadav R.K.S., Sagar R.: 2004, MNRAS 349, 1481
 
\bibitem[York et al.(2000)]{York00} York D.G., Adelman J., Anderson J.E.Jr., Anderson S.F., Annis J., Bahcall N.A., Bakken J.A., Barkhouser R., Bastian S., Berman E., and 134 coauthors: 2000, AJ 120, 1579
\end{thebibliography}
\end{document}